\documentclass[aps,showpacs,preprintnumbers,amsmath,amssymb,pra]{revtex4}

\usepackage{graphicx}
\usepackage{dcolumn}
\usepackage{bm}
\usepackage{color}

\begin{document}

\title{Collision between a dark soliton and a linear wave in an optical fiber}

\author{T. Marest,$^{1}$ C. Mas Arab\'{i},$^1$ M. Conforti,$^1$ A. Mussot,$^1$ C. Mili\'{a}n,$^2$ D.V. Skryabin,$^{2,3}$ and A. Kudlinski$^{1,}$}
\email{Corresponding author: alexandre.kudlinski@univ-lille1.fr}

\affiliation{
$^1$Univ. Lille, CNRS, UMR 8523 - PhLAM - Physique des Lasers Atomes et Mol\'{e}cules, F-59000 Lille, France\\
$^2$Department of Physics, University of Bath, Bath, United Kingdom\\
$^3$ITMO University, Kronverksky Avenue 49, St. Petersburg 197101, Russian Federation}

\begin{abstract}
We report an experimental observation of the collision between a linear wave propagating in the anomalous dispersion region of an optical fiber and a dark soliton located in the normal dispersion region. This interaction results in the emission of a new frequency component whose wavelength can be predicted using phase-matching arguments. The measured efficiency of this process shows a strong dependency with the soliton grayness and the linear wave wavelength, and is in a good agreement with theory and numerical simulations.
\end{abstract}
													

\maketitle

\section{Introduction}

When a bright soliton and a linear wave co-propagate in an optical fiber, four-wave mixing (FWM) interactions can occur and give rise to new frequency components \cite{Yulin2004,Skryabin2005,Efimov2005,Skryabin2010}. This nonlinear interaction can be physically seen as a frequency shift of the linear "probe" wave which reflects partially from the bright soliton. To study this interaction, the linear wave can be launched into the fiber directly with the soliton, as in the case of \cite{Yulin2013a,Gu2015}. The linear wave can also be emitted directly by the soliton itself and is interpreted as dispersive Cherenkov radiation \cite{Wang2015}. In this case, the dispersive wave is emitted by the bright soliton under the action of higher-order dispersion when its spectrum leaks across the zero-dispersion wavelength (ZDW) \cite{Wai1987,Akhmediev1995}. The soliton then decelerates due to the Raman effect which allows the collision between the emitted dispersive wave and the soliton \cite{Skryabin2010}. Due to its crucial importance in the supercontinuum generation process, this interaction has been the subject of intense research over the last decade \cite{Skryabin2010}. These studies have shown that the FWM process results in the generation of new components at shorter wavelengths, thus explaining the spectral broadening of the supercontinuum towards the blue/UV region \cite{Gorbach2006,Skryabin2010}. This FWM process can also be interpreted in terms of an event horizon in analogy with black holes \cite{Philbin2008,Webb2014,Wang2015}.

The interaction of bright solitons with linear waves has also been studied in the case of high-order solitons \cite{Oreshnikov2015a}, orthogonally polarized solitons \cite{Mas2016}, or in cavities made by solitons in which multiple collisions can be observed \cite{Yulin2013a,Driben2013,Wang2015optlett,Voytova2016}. 
The collision of linear waves with dark solitons has been also investigated theoretically and numerically in \cite{Oreshnikov2015}. In this work, it has been demonstrated that a linear wave propagating on top of a black soliton background may undergo the reflection process described above leading to the frequency shift of the probe wave. However, this process has not been observed experimentally yet with any dark solitons.

In this work, we report the experimental observation of the collision between a dark soliton (black or gray) with a weak linear wave. In agreement with theoretical analysis and numerical simulations, we show that this interaction leads to the generation of a new frequency component whose conversion efficiency varies with the dark soliton grayness as well as the linear wave frequency.

\section{Theory}

\subsection{Dark Soliton}
The evolution of the field with a slowly varying envelope $A(z,t)$ propagating in an optical fiber can be described by the following generalized nonlinear Schr\"{o}dinger equation (NLSE):
\begin{equation}
{i\partial_zA+D(i\partial_t)A+\gamma |A|^2A=0}
\label{GNLSE}
\end{equation}
where $D(i\partial_t)=\sum_{n \geq 2} \frac{\beta_n}{n!}(i\partial_t)^n$ is the dispersion operator expanded around $\omega_0$, the carrier frequency. $t$ is the retarded time in the frame travelling at the group velocity $V_0=V_0(\omega_0)=\beta_1^{-1}$ of the soliton carrier frequency and $\gamma$ is the nonlinear parameter. We checked that Raman and self-steepening effects do not play any significant role in the present study, thus they will be neglected.

If we only consider normal second-order dispersion ($n = 2$ and $\beta_2 > 0$), Eq. \eqref{GNLSE} admits a dark soliton solution $A_{DS}(z,t)$ of the form \cite{Kivshar98}
\begin{equation}
A_{DS}(z,t)=\sqrt{P_0}\left(\cos{\varphi} \tanh{\left(\frac{t-\sqrt{\gamma \beta_2 P_0}z}{T_{0}} \cos{\varphi}\right)}-i \sin{\varphi}\right) e^{i\gamma P_0 z}
\label{DS}
\end{equation}
where $T_{0}$ and $P_0$ are respectively the duration and peak power of the dark soliton. $\varphi \in [-\pi/2,\pi/2]$ is the grayness parameter of the soliton, $\varphi=0$ corresponds to the so-called black soliton and $\varphi\neq 0$ to gray solitons. Dark solitons are also associated with a phase jump localized at the center of the dip. A black soliton has an abrupt $\pi$ phase jump while gray solitons have smaller and smoother phase jumps. Furthermore, it has to be noted that when $\varphi < 0$ (resp. $\varphi > 0$) the soliton travels faster (resp. slower) than the carrier frequency.

In presence of higher-order dispersion ($n > 2$), dark solitons are perturbed and are known to emit radiations \cite{Karpman1993,Afanasjev1996,Milian2009} similarly to bright solitons. The dispersive wave emission from dark solitons was unambiguously observed very recently \cite{Marest2016,Marest2018}. Here, we focus on the interaction occurring when a dark soliton and a linear (dispersive) wave collide in the presence of significant higher-order dispersion. In the following, we limit our investigations to third-order dispersion term $\beta_3$ ($n = 3$) and we neglect higher-order terms.


\subsection{Phase-matching}

The interactions between linear waves and bright solitons has been theoretically studied in \cite{Yulin2004,Skryabin2005,Efimov2005,Skryabin2010,Choudhary2012}. In these works, it has been shown that the FWM interaction between linear waves and bright solitons leads to the generation of new spectral components. This process follows a phase matching relation, involving soliton and linear wave properties, that can be derived by a perturbative approach \cite{Akhmediev1995}. In a similar way, Oresnikov \emph{et al. } \cite{Oreshnikov2015} used this perturbative method to derive the phase matching condition in the specific case of a linear wave interacting with a black soliton. In the present study, we generalize this approach to take into account the grayness of the dark soliton interacting with a linear wave. A similar phase marching condition has been obtained for the radiation emitted by shock waves \cite{PRA2013,OL2014,SciRep2015}.  In the following, the linear wave is externally launched at the fiber input and therefore acts as a probe. It will therefore be termed probe wave, following the terminology used in the literature.

The dispersion relation of linear waves propagating over a background is (see Appendix \ref{App:phase_match})
\begin{equation}
k(\Omega)=-\beta_{1sol}\Omega+\frac{\beta_2}{2}\Omega^2+\frac{\beta_3}{6}\Omega^3+k_{NL}
\label{kx}
\end{equation}
where $\Omega=\omega-\omega_0$ and $\omega$ is the frequency of the linear wave. Three main contributions are involved in this relation : the dark soliton (first term), in which $\beta_{1sol}=\sqrt{\gamma \beta_2 P_0}\sin(\varphi)$ is the soliton group velocity, the dispersion of linear waves (second and third terms) and a nonlinear contribution $k_{NL}=\gamma P_0$ due to the background of the dark soliton (last term). The resonance condition leading to the new frequency generation is verified when the wave vectors of the probe and generated waves are equal, which leads to the following equation

\begin{equation}
-\beta_{1sol}\Omega+\frac{\beta_2}{2}\Omega^2+\frac{\beta_3}{6}\Omega^3+k_{NL} = k(\Omega_p)
\label{ph_match}
\end{equation}
where $\Omega_p=\omega_p-\omega_0$ and $\omega_p$ is the absolute frequency of the probe wave. The real solution of Eq. \eqref{ph_match} gives the frequency of the generated wave. Note that in the case of a black soliton ($\varphi=0$ and therefore $\beta_{1sol}=0$), the expression is the same than for a bright soliton \cite{Skryabin2005}. However, when $\varphi \neq 0$, the soliton carries an extra momentum whose effect is to change the frequency of the generated wave. Equation \eqref{ph_match} also shows that the collision process depends on the fiber properties ($\beta_2$, $\beta_3$ and $\gamma$) as well as the probe wave frequency ($\omega_p$). In the following, the frequencies predicted by Eq. \eqref{ph_match} will be compared with the experimental and simulation results.


\subsection{Conversion efficiency}

In order to fully describe the collision process, we focus now on the amplitude of the generated wave. If we refer to the case of bright solitons \cite{Choudhary2012,Wang2015}, this amplitude can be calculated by considering the soliton as a potential. In this case, the probe and generated signal waves are the transmitted and reflected waves, respectively. However, in the case of quasi-continuous linear wave, only a small part of the probe interacts with the soliton. The duration of this interaction can be approximated as $T_{in}=L_{fiber}/(\Delta V_{sp})$, where $\Delta V_{sp}$ is the group velocity mismatch between the dark soliton and probe wave. The energy of the probe interacting with the soliton can be expressed in the time domain as $E_{in}=P_p \times T_{in}$, where $P_p$ is the probe wave peak power.

Consequently, the energy of the generated wave is taken from the reflected part of the probe wave interacting with the soliton : $E_g = R \times E_{in}$, where $R$ is the reflection coefficient corresponding to the soliton potential (see appendix \ref{App:Energie}). However, the experimental peak power of the probe wave is generally not exactly known. Therefore, we express the energy of the generated wave in the normalized form \cite{Choudhary2012}:

\begin{equation}
E_g \propto R|v_{sp}^{-1}| = R|{\beta_2}\Omega_p+\frac{\beta_3}{2}\Omega_p^2-\sqrt{\gamma \beta_2 P_0}\sin(\varphi)|
\label{energie_th}
\end{equation}
As expected, when the group velocity mismatch between the soliton and the probe wave becomes large, the energy of the generated wave tends to zero. However, as we can see, the generated wave energy mainly depends on the soliton grayness and the probe wave frequency. Therefore, the impact of those parameters will be studied experimentally in what follows.

\section{Simulation}

\begin{figure}[!htbp]
\centering
\includegraphics[width=0.95\linewidth]{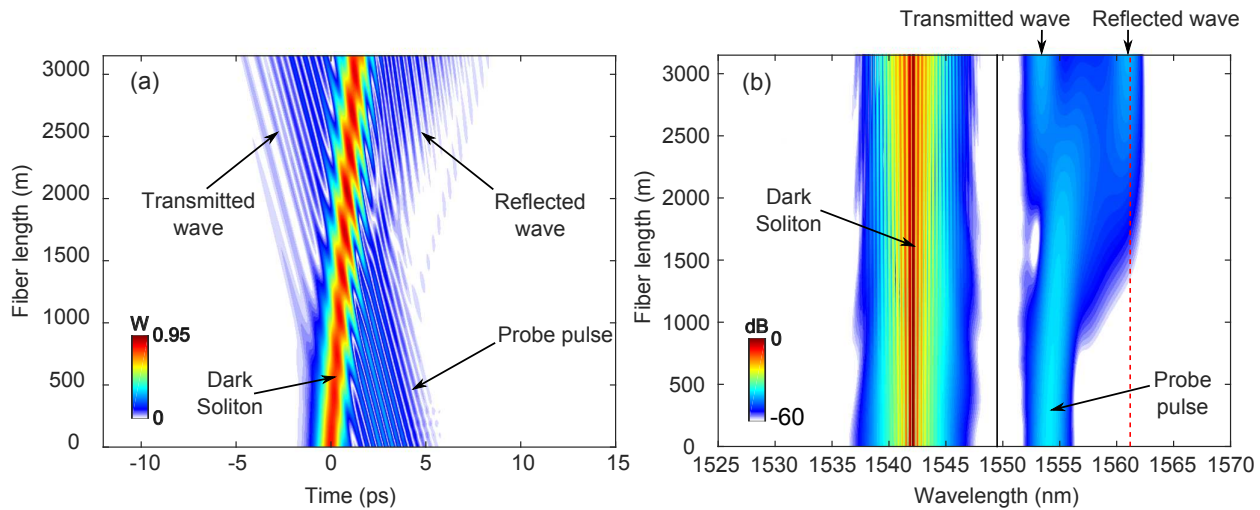}
\caption{Numerical simulation of the collision between a gray soliton and a weak probe wave. (a) temporal evolution during the propagation in the DSF. The field shown is $||A(z,t)|^2-P_0|$. (b) spectral evolution versus fiber length. The black horizontal line represents the ZDW
of the fiber and the dashed red line represents the theoretical wavelength of
the wave produced by the collision [solution of Eq. \eqref{ph_match}].}
\label{fig:simu}
\end{figure}

To illustrate the collision process between a dark soliton and a probe wave, we performed numerical simulations based on Eq. \eqref{GNLSE}. Thus, we take the example of a gray soliton of grayness $\varphi=\pi/8$, duration $T_0 = 600$ fs and peak power $P_0=0.94$ W colliding with a weak Gaussian probe wave $A_p(t)$ of the following form:

\begin{equation}
A_{P}(t)=\sqrt{P_p}\exp{\bigg[-\left(\dfrac{t-\tau}{\sqrt{2}T_{P}}\right)^2\bigg]} e^{-i(\omega_p-\omega_0)t}
\label{HG}
\end{equation}
where $P_p = 0.02 \times P_0 = 18.9$ mW. The full-width at half maximum (FWHM) duration of the probe pulse ($T_{FWHM} = 1.665~T_p$) is equal to 2 ps. The delay between the probe wave and the soliton is set to $\tau=3$ ps. Figures \ref{fig:simu}(a) and (b) represents respectively the temporal and spectral evolution of the collision process in a 3.15 km long dispersion shifted fiber (DSF) whose ZDW is located at 1549.5 nm. The soliton and probe wavelengths are respectively $\lambda_0=1542$ nm and $\lambda_p=1554$ nm, corresponding to $\beta_2=0.68$ ps$^2$/km and $\beta_{2}=-0.405$ ps$^2$/km, respectively. The nonlinear parameter $\gamma$ is equal to 2~W$^{-1}$.km$^{-1}$ and the third order dispersion $\beta_3$ is equal to 0.115 ps$^3/$km. Note that the temporal dynamics in Fig. \ref{fig:simu} (a) represents the field intensity minus the soliton background power $||A(z,t)|^2-P_0|$, for the sake of clarity.


As can be seen in Fig. \ref{fig:simu}(a), the probe wave and the soliton collide around 1 km of propagation. After the collision, a part of the probe is transmitted and the other one is reflected. In the spectral domain, one can see that the probe and the transmitted wave wavelengths are equal. Around 2 km of propagation a new spectral component associated to the reflected wave is emitted at 1561 nm. This generated wave can be attributed to the FWM interaction between the dark soliton and the probe wave occurring at their collision and resulting in the frequency shift of the reflected wave. The theoretical wavelength of the generated wave can be calculated from the phase matching condition given by Eq. \eqref{ph_match} and is drawn in red dotted line in Fig. \ref{fig:simu}(b). It corresponds to 1561 nm, which is in very good agreement with the simulation and therefore confirms the validity of the previous analysis.

\section{Experimental setup}

\begin{figure}[!htbp]
\centering
\includegraphics[width=0.90\textwidth]{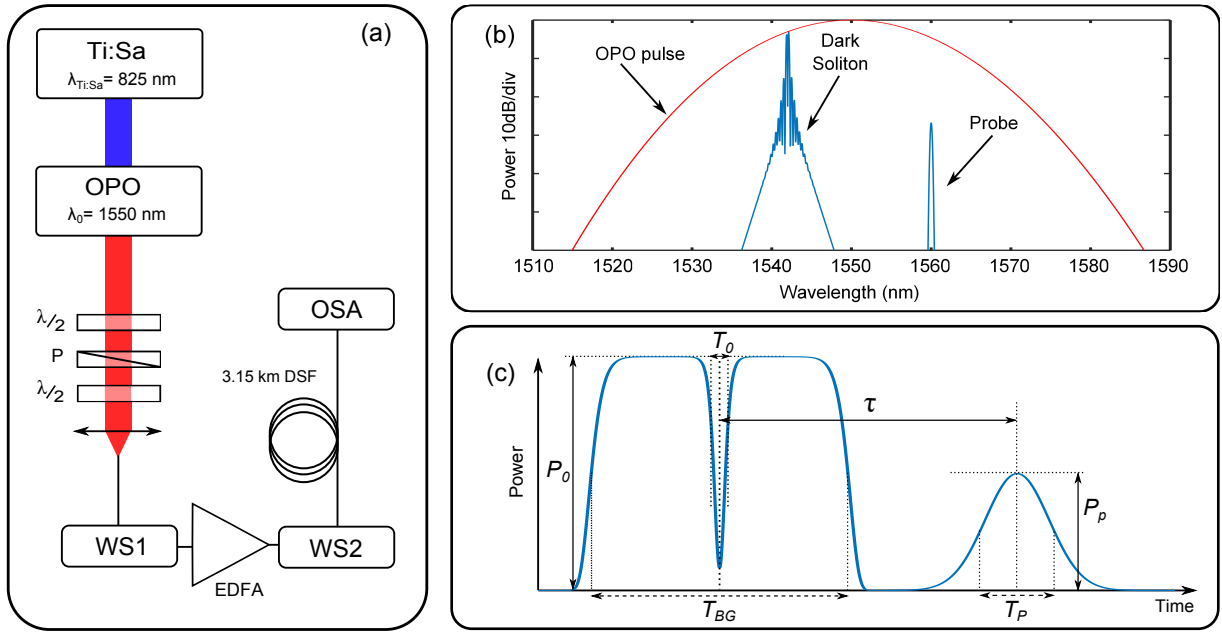}
\caption{(a) Experimental setup. P: Polarizer; WS: waveshaper; $\lambda/2$: half-wave plate; EDFA: erbium-doped fiber amplifier; OSA: optical spectrum analyzer. (b) Schematic spectrum and (c) time profile of the total field launched into the fiber. }
\label{fig:setup}
\end{figure}

In order to experimentally study the collision process, we need to generate both a dark pulse (required to excite a dark soliton in the fiber) and a probe pulse, but we also need tp control the soliton grayness as well as the time delay between the two pulses. Several methods have been proposed to generate dark solitons \cite{Emplit1987,Kroekel1988,Weiner1988,Hong1991,Rothenberg1992,Emplit1992,Emplit1997,Cao1999,Finot2006,Song2014}. However, none of them allows an accurate control of the soliton grayness. To do that, we implemented a technique based on phase and spectral shaping using commercial waveshapers (Finishar wavesapher 4000S), inspired of \cite{Marest2018}. The corresponding experimental setup, shown in Fig. \ref{fig:setup}(a), will be used to generate simultaneously a dark pulse $A_{DS}(0,t)$ located on a super-Gaussian background pulse $A_{BG}(t)$ and a Gaussian pulse $A_p(t)$ which will act as the probe wave. The total field launched in the fiber can be written $A(t)=A_{DS}(0,t) \times A_{BG}(t)+A_{P}(t)$ with
\begin{equation}
A_{BG}(t)=\exp{\bigg[-\left(\dfrac{t}{T_{BG}}\right)^n\bigg]}
\label{HG}
\end{equation}
where $T_{BG}$ is the duration of the super-Gaussian background pulse and $n$ its order.

In order to synthesize this field, we use an optical parametric oscillator (OPO) pumped by a Ti:Sa laser. It delivers Gaussian pulses of 220 fs FWHM duration and tunable around 1550 nm at 80 MHz. The pulses pass though a variable attenuator consisting of half-wave plates and a polarizer and reaches the first waveshaper WS1. Here, the initial pulse is shaped in amplitude (without shaping the  phase jump characteristics of dark solitons and delay between the two pulses). This step is represented in Fig. \ref{fig:setup}(b) and (c) in which the large OPO spectrum [red curve] is shaped into a dark pulse of duration $T_0$ on a wide super-Gaussian background of duration $T_{BG}$ and a probe wave of power and duration $P_p$, $T_p$ [blue curve]. After WS1, this field is amplified with an erbium doped fiber amplifier (EDFA) and reaches the second waveshaper WS2. At this stage, the odd-symmetry phase profile of the dark pulse is applied and the time delay between the dark pulse and the probe pulse is fixed thanks to a linear phase variation of the probe. Additionally, the use of this second waveshaper allows to filter out the amplified spontaneous  emission added during the amplification and to correct spectral distortions due to amplification. For all simulations and experiments below, we use the DSF described in the previous section. The parameters of the dark pulse are set to: $\lambda_0=1542$ nm ($\beta_{2}=0.6795$ ps$^2/$km), $P_0=0.94$ W, $T_{0}=600$ fs, $T_{BG}=40 \times T_{0}=24$ ps, $n=16$. 

To summarize, with the experimental setup presented in Fig. \ref{fig:setup}(a), it is possible to generate a dark pulse located onto a super-Gaussian background pulse with a controlled grayness and wavelength, co-propagating with a linear probe pulse with a controlled delay, duration, wavelength and peak power.

\section{Impact of the dark soliton grayness}
\begin{figure}[!htbp]
\centering
\includegraphics[width=0.85\linewidth]{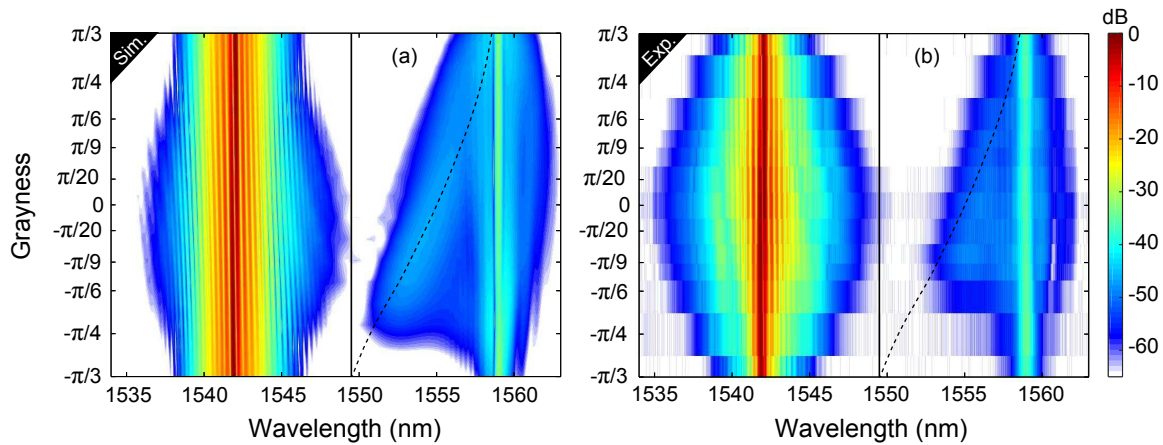}
\caption{(a) Simulated and (b) experimental fiber output spectrum against the grayness of the dark soliton. The black horizontal line represents the ZDW of the fiber. The dashed black lines stand for the theoretical wavelength of the generated wave produced by the collision [obtained from Eq. \eqref{ph_match}].}
\label{gray}
\end{figure}

In this section, we study the impact of dark soliton grayness on the collision process. Figure \ref{gray}(a) represents the simulated fiber output spectrum against the soliton grayness changed from $\varphi = -\pi/3$ to $\varphi=\pi/3$. The probe wavelength is fixed to a constant value of $\lambda_p = 1559$ nm ($\beta_{2p}=-0.852$ ps$^{2}$/km) with a FWHM duration of 5.9 ps. Its peak power is fixed to $P_p=0.0035 \times P_0=3.3$ mW and the time delay to $\tau= 6.5$ ps. The black solid line represents the fiber ZDW. As can be seen, the generation of the new spectral component in the anomalous dispersion region occurs for grayness values from $-\pi/6$ to $\pi/4$. Its wavelength increases with the grayness  because the carrier wavelength of the dark solitons shifts with grayness. As in the previous section, the theoretical wavelength of the reflected wave can be found from the solution of Eq. \eqref{ph_match} for each grayness value. The theoretical evolution of this wavelength is represented in black dashed line in Fig. \ref{gray}. The agreement between theoretical results and numerical simulation is very good.


To observe this dynamics experimentally, the method shown in Fig. \ref{fig:setup} is used with the same parameters. The corresponding results are reported in Fig. \ref{gray}(b) which represents the evolution of the fiber output spectrum as a function of the grayness. In agreement with the simulations results, the generated wave wavelength increases for grayness values from $\varphi = -\pi/8$ to $\varphi = +\pi/3$, following the same tendency as the theoretical prevision [black dashed line]. However, when the grayness is below $\varphi = -\pi/6$, one can notice the absence of emission. The small discrepancy between simulations and experiments can be attributed to experimental uncertainties. Figure \ref{gray} also shows that the generated wave observed experimentally is efficiently emitted for a specific range of grayness values. 
Hereafter, we will focus on the efficiency of the collision process.

\begin{figure}[!htbp]
\centering
\includegraphics[width=7.5cm]{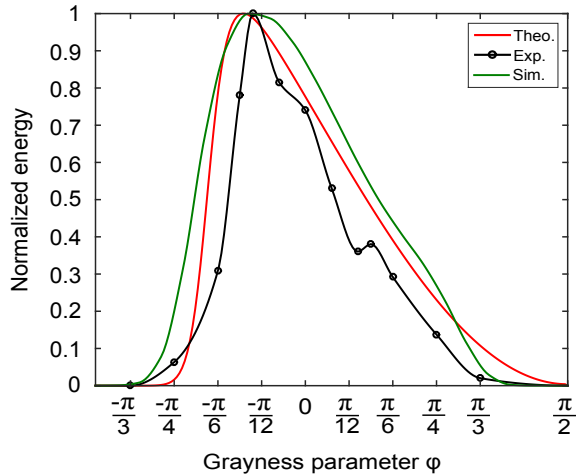}
\caption{Energy of the generated wave versus the soliton grayness. The red and green curves correspond respectively to the theoretical prediction [Eq. \eqref{energie_th}] and numerical results. Black circles are experimental measurements. The black line is a guide for the eye. }
\label{energie_phi}
\end{figure}

Figure \ref{energie_phi} presents the evolution of the generated wave energy as a function of the soliton grayness. The theoretical result, obtained from Eq. \eqref{energie_th}, is represented in red. The simulation and experimental results (respectively shown in green and black) are calculated by integrating the power spectral density over the emission spectral range and dividing by the repetition rate of the laser. Theoretical predictions, simulations and experiments are in good agreement and show that the efficiency of the collision process presents a maximum for grayness value around $-\pi/8$. When the grayness tends to $-\pi/3$, the generated wave energy decreases rapidly to become almost zero before $-\pi/4$. The same happens for increasing grayness values, \emph{i.e.} the generated wave energy decreases and almost vanishes for grayness values larger than $\pi/3$.

\section{Impact of the probe wavelength}
To study the impact of probe frequency on the collision process, the soliton grayness is now fixed to a constant value of $\varphi = 0$ (black soliton case). The wavelength of the probe is varied from 1552 nm to 1564 nm with the waveshapers, corresponding to $\beta_2$ values varying between -1.297 and -0.225 ps$^2$/km. Starting with numerical simulations, the peak power of the probe wave is set to $P_p = 0.005 \times P_0 = 4.7$ mW and the FWHM duration of the probe to 16 ps with a zero time delay between the soliton and the probe.
\begin{figure}[!tbp]
\centering
\includegraphics[width=0.85\linewidth]{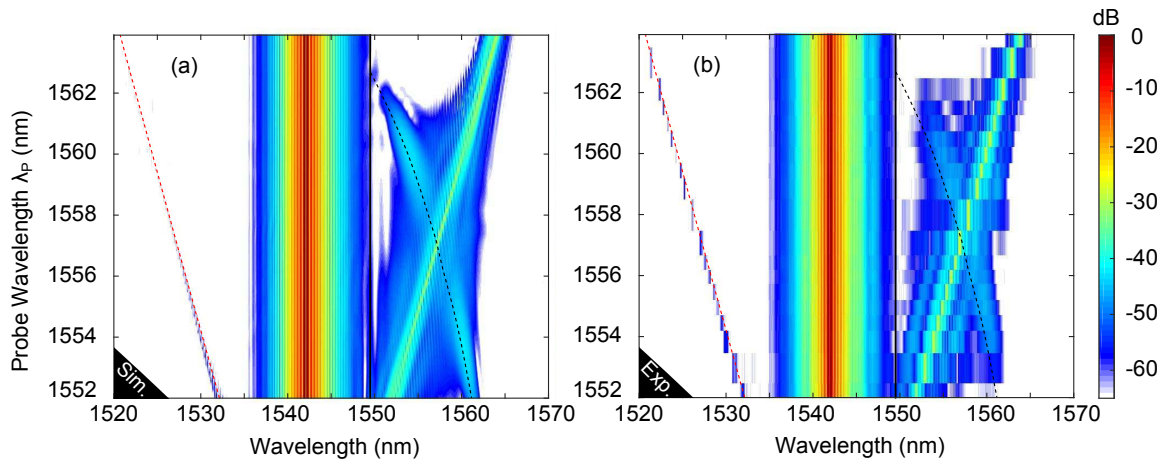}
\caption{(a) Simulated and (b) experimental fiber output spectrum versus the probe wave wavelength. The black horizontal line represents the ZDW of the fiber. The dashed black lines represent the theoretical wavelength of the wave produced by the collision between a black soliton ($\varphi = 0$) and the probe wave [obtained from Eq. \eqref{ph_match}]. The red lines correspond to the FWM interaction between the probe and the soliton background (see text).}
\label{lambda}
\end{figure}
Figure \ref{lambda}(a) represents the evolution of the simulated output spectrum as a function of the probe frequency. It can be seen that the generated wave wavelength decreases from 1561 nm to 1550 nm when the probe wavelength increases from 1551 nm to 1562 nm.  This results is in very good agreement with the solution of  the phase matching condition [black dashed line in Fig. \ref{lambda}], obtained from Eq. \eqref{ph_match}.
One can also note the presence of a spectral component on the left of the spectrum which varies from 1532 nm to 1521.4 nm with the probe frequency. This peak follows the red dashed line that corresponds to the FWM process between the soliton background and the probe wave at $2\omega_0-\omega_p$ (the second peak located at long wavelength $2\omega_p-\omega_0$ being too weak to be seen experimentally).

Figure \ref{lambda}(b) shows the experimental results obtained using the same parameters as in the simulation. One can see that the generated wave wavelength follows the same evolution as in the simulation. Therefore, this result is in very good agreement with both the experimental and theoretical results and confirms the previous analysis. 

\begin{figure}[!bp]
\centering
\includegraphics[width=7.5cm]{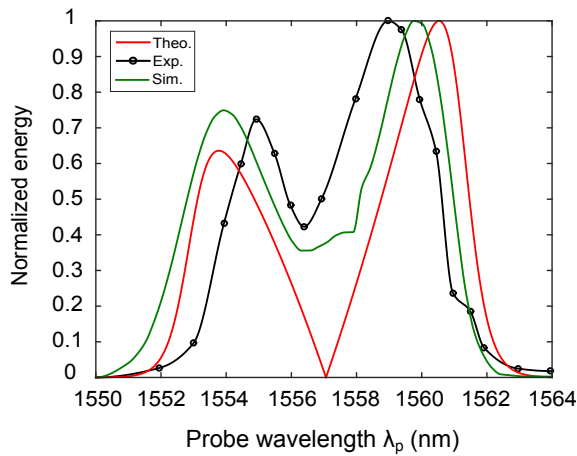}
\caption{Energy of the generated wave versus the probe wavelength. The red and green curves correspond respectively to the theoretical prediction [Eq. \eqref{energie_th}] and numerical results. Black circles are experimental measurements. the black line is a guide for the eye. }
\label{energie_lambda}
\end{figure}

Figure \ref{energie_lambda} shows the theoretical evolution of the energy of the generated wave (red curve),  based on Eq. \eqref{energie_th}. We can see two local maxima, the first one being at 1560.5 nm and the second one at 1553.8 nm. The generated wave energy falls rapidly to zero for values less than 1551 nm and greater than 1563 nm. Furthermore, when the probe wavelength tends to 1557 nm, the group velocity of the probe and the soliton are almost equal, implying 	a vanishing theoretical energy of the generated wave, as it is evident from Eq. (\ref{energie_th}).
To analyze the evolution of the generated wave energy, we followed the same integration method as before. Figure \ref{energie_lambda} shows the experimental and numerical energy of the generated wave, in black and green respectively. The agreement with theory is qualitatively good, \emph{i.e.} two local maxima can also be found. One can note that the energy does not fall to zero around 1557 nm as theoretically predicted. This discrepancy with the theoretical curve can be attributed to the fact that the probe wave is not a continuous wave and therefore has a relatively broad linewidth. Indeed, the spectral width of the probe directly impacts the resolution of the energy measurement. In addition, the spectral width may increase during propagation due to cross-phase modulation with the soliton background. Thus, the zero of the theoretical curve, seen around 1557 nm, cannot be found in this case. However, when the probe wave wavelength increases ($>1562$ nm), its group velocity mismatch with the soliton increases too, which strongly reduces the conversion efficiency, as in the theoretical evolution. The same arguments apply for decreasing wavelengths ($<1551$ nm), showing that the group velocity mismatch between the dark soliton and the probe wave is the control parameter of the new frequency generation and of its efficiency. Despite the quantitative discrepancies, these results confirm that the collision between a black soliton and a linear wave follows the same phase matching rule as in the previous case where the soliton grayness was changed during the experiments.


\section{Conclusion}
To conclude, in this work we made the first experimental observation of the collision between a dark soliton (gray and/or black) and a weak linear wave. By generating both the dark soliton and the linear wave with available commercial waveshapers, it has been possible to control all the parameters of both the dark soliton and the linear wave. It has been demonstrated that the collision process leads to the generation of a new frequency components whose conversion efficiency varies with the dark soliton grayness as well as the linear wave frequency, which are all in good agreement with theoretical and numerical predictions.

\section*{Funding}
This work was partly supported by IRCICA, CNRS, USR 3380, by the NoAWE (ANR-14-ACHN-0014) project, by the "Fonds Européen de Développement Economique Régional", the Labex CEMPI  (ANR-11-LABX-0007)  and  Equipex  FLUX  (ANR-11-EQPX-0017) through the "Programme Investissements d'Avenir". We also acknowledge funding from the CNRS and RFBR (16-52-150006) through the Russian-French PRC program.  D.V.S. acknowledges support through the ITMO visiting professorship scheme.

\appendix

\section{Appendix}
\subsection{Derivation of phase-matching relation}
\label{App:phase_match}

In this appendix, we present the detailed calculation of the phase-matching condition corresponding to the four wave mixing process occurring at a collision between a dark soliton and a linear wave. Starting from Eq. \eqref{GNLSE}:
\begin{equation}
i\partial_zA+D(i\partial_t)A+\gamma |A|^2A=0
\label{eq:NLS}
\end{equation}
we write the field envelope $A(z,t)$ as a superposition of a dark soliton, ($u_{DS}(z,t) e^{i\gamma P_0 z}$) given by Eq. \eqref{DS}, and a weak wave ($g(z,t)$) \cite{Skryabin2005}:
\begin{equation}
A(z,t)=(u_{DS}(t,z)+g(z,t))\text{e}^{i\gamma P_0 z}.
\end{equation}
By substituting $A(z,t)$ in \eqref{eq:NLS} and linearizing $g(z,t)$, we find the equation describing the evolution of $g(z,t)$:
\begin{equation}
i\partial_zg+D(i\partial_t)g+\gamma(2|u_{DS}(z,t)|^2-P_0)g+u_{DS}(z,t)^2g^*)=0
\label{eq:pertorbacio_general}
\end{equation}
By moving to a reference frame where the soliton is at rest ($\tau=t-\sqrt{\gamma \beta_2 P_0}z$) this equation can be written :
\begin{equation}
i\partial_zg+\bar{D}(i\partial_\tau)g+\gamma((2|u_{DS}(\tau)|^2-P_0)g+u_{DS}(\tau)^2g^*)=0 \quad,\quad \bar{D}(i\partial_\tau)=-i\sqrt{\gamma \beta_2 P_0}\partial_\tau+D(i\partial_\tau)\label{eq:pertorbacio}
\end{equation}
Now, we consider $g(z,\tau)$ to be the sum of the probe and a wave $\psi$ generated elsewhere:
\begin{equation}
g(z,\tau)=w\text{e}^{i(k (\Omega_p)z-\Omega_p\tau)}+\psi(z)\text{e}^{i(k (\Omega)z-\Omega\tau)}
\end{equation}
where $k (\Omega_p)$ is the dispersion relation of the probe wave. The wave $\psi$ can be efficiently amplified if one of the following conditions is satisfied \cite{Skryabin2005,Oreshnikov2015,OL2014,SciRep2015}:
\begin{equation}
k(\Omega)=\pm k(\Omega_p)
\end{equation}
In the experiments and simulations, only the case where $k(\Omega)=k(\Omega_p)$ is observed. The dispersion relation of linear waves propagating  over a continuous background on the reference frame of the soliton is given by:

\begin{equation}
k(\Omega)=-\sqrt{\gamma \beta_2 P_0}\sin(\varphi)\Omega+\frac{\beta_3}{6}\Omega^3+\frac{\Omega}{2}\sqrt{\beta_2(\beta_2\Omega^2+4\gamma P_0)}
\label{kx_origine}
\end{equation}
In the low power regime ($|\beta_2 \Omega^2|>>|\gamma P_0|$), this expression can be simplified to [Eq. \eqref{kx}]) :
\begin{equation}
k(\Omega)=-\sqrt{\gamma \beta_2 P_0}\sin(\varphi)\Omega+\frac{\beta_2}{2}\Omega^2+\frac{\beta_3}{6}\Omega^3+\gamma P_0
\end{equation}

\subsection{Calculation of generated wave amplitude}
\label{App:Energie}
Phase-matching arguments do not provide any information about the process efficiency. To find out this, we consider the interaction between the probe and the dark soliton as a scattering problem \cite{Choudhary2012}. In this picture, the generated wave (probe) plays the role of reflected (transmitted) wave and the soliton is seen as the scattering potential.
Taking advantage of the phase-matching condition, the perturbation can be expressed as $g=\Psi(\tau)e^{ikz}$ and substituted in equation \ref{eq:pertorbacio}. Dropping non-phase matched terms and making the approximation $k\approx \bar{D}(\Omega)+\gamma P_0$, Eq. \eqref{eq:pertorbacio} becomes :
\begin{equation}
\bar{D}(i\partial_\tau)\Psi-(2q\cos^2(\varphi)\text{sech}^2(\tau\cos(\varphi)/t_0)g+\bar{D}(\Omega))\Psi=0
\end{equation}
The studied interaction takes place in the neighbourhood of the frequency which is group-velocity matched ($\Omega_{GVM}$) with the soliton. For this reason, the problem can be simplified by performing a phase-rotation such as $\tilde{\Psi}=\Psi e^{i\Omega_{GVM}\tau}$:

\begin{equation}
\left(-\frac{|\beta_2(\Omega_{GVM})|}{2}\frac{d^2}{d\tau^2}+(2q\cos^2(\varphi)\text{sech}^2(\tau\cos(\varphi)/t_0)-\Delta D)\right)\Psi=0,
\label{eq:SE}
\end{equation}
where $\Delta D=-\bar{D}(\Omega)+\bar{D}(\Omega_{GVM})$, $\beta_2(\Omega_{GVM})=\beta_2+\beta_3 \Omega_{GVM}$ and $\Omega_{GVM}$ can be expressed as:
\begin{equation}
\Omega_{GVM}=\frac{-\beta_2-\sqrt{\beta_2^2+2\sin(\varphi)\beta_3\sqrt{\gamma \beta_2 P_0}}}{\beta_3}.
\end{equation}
Note that higher-order dispersion terms have been neglected. This approximation can be safely performed because near $\Omega_{GVM}$, the dispersion can be considered as parabolic. The reflection of the potential in Eq. \eqref{eq:SE} is \cite{Landau1977}:
\begin{equation}
R=\frac{\cosh^2\left(\frac{\pi}{2}\sqrt{16\frac{|\beta_2 (\Omega_{G})|}{\beta_2}-1} \right)}{\cosh^2\left(\frac{\pi}{2}\sqrt{16\frac{|\beta_2 (\Omega_{G})|}{\beta_2}-1} \right)+\sinh^2\left(\frac{\pi t_0 \Delta\Omega}{cos(\varphi)}\right)}
\label{reflex}
\end{equation}
where $\Delta\Omega=\Omega_p-\Omega_{GVM}$.

It is important to note that when $\varphi=0$ (black soliton case), Eq. \eqref{reflex} is the same as in the case of a bright fundamental soliton colliding with a linear wave \cite{Wang2015}. Thus the reflection coefficient $R$ does not depend on $\beta_2$. However, when $\varphi \neq 0$ (gray soliton), the reflection coefficient $R$ depends on $\beta_2$. Equation \eqref{reflex} also shows that the total conversion efficiency ($R=1$) is found when the probe and the dark soliton are group velocity matched ($\Delta \Omega=0$).

\end{document}